\documentclass[aps,prb,reprint,superscriptaddress,amsmath,amssymb,floatfix]{revtex4-2}

\usepackage{graphicx}
\usepackage{dcolumn}
\usepackage{bm}
\usepackage{xcolor}
\usepackage{booktabs}
\usepackage[colorlinks=true,linkcolor=blue!60!black,citecolor=blue!60!black,urlcolor=blue!60!black]{hyperref}

\begin{document}

\title{Chiral-Nematic Lattice Instability in Charge Density Wave of 1$T$-TiSe$_2$}

\author{Abdurrafi Ahsan Alghifari}
\affiliation{Center for Artificial Low Dimensional Electronic Systems, Institute for Basic Science (IBS), Pohang 37673, Republic of Korea}
\affiliation{Department of Physics, Pohang University of Science and Technology (POSTECH), Pohang 37673, Republic of Korea}

\author{Euihwan Do}
\affiliation{Center for Artificial Low Dimensional Electronic Systems, Institute for Basic Science (IBS), Pohang 37673, Republic of Korea}

\author{Jae Whan Park}
\affiliation{Center for Artificial Low Dimensional Electronic Systems, Institute for Basic Science (IBS), Pohang 37673, Republic of Korea}

\author{Han Woong Yeom}
\email{yeom@postech.ac.kr}
\affiliation{Center for Artificial Low Dimensional Electronic Systems, Institute for Basic Science (IBS), Pohang 37673, Republic of Korea}
\affiliation{Department of Physics, Pohang University of Science and Technology (POSTECH), Pohang 37673, Republic of Korea}

\date{\today}

\begin{abstract}
Chiral charge-density waves (CDWs) are a rare form of electronic order with handedness, enabling novel couplings among light, lattice, and topology in quantum materials. Yet the microscopic origin of chirality has remained elusive, even in the prototypical chiral CDW system 1$T$-TiSe$_2$. Using \textit{ab initio} calculations, we investigate the structural stability and electronic structure of competing CDW models in 1$T$-TiSe$_2$. We show that a chiral monoclinic ($C2$) structure is stabilized by the cooperative condensation of multiple phonon modes and reproduces the chiral topography and energy gap observed experimentally. 
We further find an exceptionally soft potential landscape around this energy-minimized structure, along the direction that removes the remaining rotational symmetry, allowing the lattice to fluctuate through fully asymmetric configurations while its average structure remains chiral. This dynamical chiral CDW naturally reconciles the chiral domains seen by time-averaged topographic probes with the complete loss of symmetry detected by bulk spectroscopy, and clarifies how chiral CDWs emerge from the interplay of lattice instability and fluctuation.
\end{abstract}

\maketitle


\section{Introduction}\label{sec1}

Charge density waves (CDWs) are a fundamental way that electrons and lattices couple to lower the total energy of a system through spontaneously breaking its translational symmetry~\cite{Gruener1994,Manzeli20172d}. 
CDWs are also important because they are often related to other exotic electronic properties as in the case of their competition or coexistence with superconductivity~\cite{Morosan2006,Kusmartseva2009,Joe2014}.
Recent progress in the investigation of CDWs has revealed extra symmetry breaking within CDW states, most notably, the formation of chiral CDW phases with broken mirror and inversion symmetry~\cite{Ishioka2010,vanWezel2011,iavarone2012,Castellan2013,Zenker2013,Gradhand2015}.
Chiral CDWs can couple to topological degrees of freedom~\cite{Li2022}, chiral light polarizations~\cite{xu20,qiu2025}, and possibly to chiral phonons~\cite{Zhang2024}. The helicity-dependent dynamics of the chiral CDW is presently under active debate~\cite{qiu2025,QiuComment2026,QiuReply2026}.
This phenomenon has recently been observed in a wider class of materials such as Kagome superconductors~\cite{Cheng2025}, where the chiral CDW is related to the time-reversal symmetry breaking in the coexisting superconducting state.
However, the driving forces of such extra symmetry breaking have been challenging to identify in many cases.

A model system for investigating this phenomenon has been 1$T$-TiSe$_2$, where the chiral CDW was discovered first~\cite{Ishioka2010}.
1$T$-TiSe$_2$ hosts a commensurate $2\times2\times2$ CDW modulation with a transition temperature $T_c \approx 200$~K~\cite{DiSalvo1976}. Suppressing this CDW order through copper intercalation or applied pressure induces a superconducting phase that peaks near the CDW critical point~\cite{Morosan2006,Kusmartseva2009,Joe2014}. The chiral distortion in this CDW state was first identified~\cite{Ishioka2010} and established~\cite{kim2024nanolett} by scanning tunneling microscopy, and recently confirmed by the circular photogalvanic effect~\cite{xu20} and angle-resolved photoelectron spectroscopy~\cite{ghafari2017}. However, despite extensive investigation, the microscopic structural model of the chiral distortion is still elusive~\cite{Wickramaratne2022}, leaving the origin of inversion-symmetry breaking and its connection to the electronic ground state unclear.

In particular, most theoretical investigations so far share the assumption that the low-temperature phase adopts the centrosymmetric $P\bar{3}c1$ structure identified in early structural studies~\cite{woo76,DiSalvo1976}. 
This achiral phase is obviously incompatible with the observation of chiral domains in STM topography~\cite{hil18,kim2024nanolett} and the chiral response in circular photogalvanic measurements~\cite{xu20}.
Furthermore, standard DFT calculations using this model typically predict a metallic (or semimetallic) band structure, requiring the inclusion of strong excitonic interaction~\cite{cercellier2007evidence, monney2010temperature, Rossnagel2011, hellmann2012time} or on-site Coulomb interaction to reproduce the experimentally observed CDW gap of about 140 meV~\cite{hellgren2017critical,bianco2015electronic,chen2015charge}. In contrast, a few recent experimental~\cite{Buchberger2025,Fragkos2026} and theoretical~\cite{Fragkos2026Arxiv,Pashov2025} studies have strengthened the case for a lattice origin, building on earlier evidence for the interplay of excitonic and phononic contributions~\cite{Hedayat2019}.

In an effort to solve this problem, recent DFT calculations showed that a monoclinic $C2$ chiral CDW structure can emerge from the interlayer shear displacement of the $P\bar{3}c1$ structure~\cite{Subedi2022}, which matches well the chiral STM topography~\cite{kim2024microscopicoriginchiralcharge,kim2024nanolett}. Two subsequent studies have independently converged on the same chiral $C2$ arrangement, through a systematic enumeration of interlayer CDW phases~\cite{Shao2026} and through an interlayer stacking-order analysis~\cite{Carpene2026}.
On the other hand, a recent spectroscopic study combined with theoretical modeling~\cite{Kim2024} suggested that a competition between different CDW components induced by charge-lattice symmetry frustration can result in a chiral CDW distortion.
This work, however, assumed the electronic order as the primary source of the instability and did not explicitly suggest any atomic structure for the chiral distortion. 
Surprisingly, this ``charge-lattice frustration'' breaks all mirror and inversion symmetries so that the chiral CDW phase has no explicit crystal symmetry ($P1$) without a sign of the $C2$ symmetry. 
This discrepancy leaves the origin of chiral CDW in this model system elusive.

In the present work, we resolve this apparent discrepancy by mapping the full energy landscape of the competing distortion modes. Building upon prior phonon surveys that identified the $C2$ structure as a harmonic minimum~\cite{Subedi2022}, our \textit{ab initio} analysis establishes that this chiral phase is a sufficient lattice-driven model to explain most of the experimental observations. It reproduces the chiral topography, the energy gap, and the domain-wall structures~\cite{kim2024nanolett}.
Furthermore, we discover that this $C2$ ground state sits at the center of a soft nematic energy manifold, where the remaining rotational symmetry can be removed at a minimal energetic cost. The lattice fluctuates through fully asymmetric configurations while its average structure remains
chiral.
This dynamic fluctuation reconciles the experimental paradox; fast spectroscopic probes detect the instantaneous $P1$ symmetry loss, while temporally-integrating topographic probes capture the spatially averaged $C2$ chirality. Consequently, the present work provides a lattice-driven mechanism of chiral CDW without invoking static phase separation~\cite{Kim2024} or assuming a frustration between charge and lattice degrees of freedom~\cite{Kim2024}.

\section{Results}\label{sec2}
\subsection{Phonon modes and possible structure models}

\begin{table}[tbp]%
\centering 
\caption{Energetic hierarchy and symmetry classification of the structural phases derived from $M$-point ($\Gamma_5^3$) and $L$-point ($\Gamma_7^3$) phonon modes. Energies ($\Delta E$) are given in meV per supercell relative to the high-symmetry parent phase.\label{tab:energetics}}%
\begin{tabular*}{\columnwidth}{@{\extracolsep\fill}lllc@{\extracolsep\fill}}
\toprule
\textbf{Family} & \textbf{Distortion} & \textbf{Symmetry} & \textbf{$\Delta E$ (meV/cell)} \\
\midrule
Reference & - & $P\bar{3}m1$ & \hphantom{-}0.00 \\
\midrule
Pure $\Gamma_5^3$ ($M$) & $1Q_M$ & $P2/c$ & -9.7 \\
                        & $2Q_M$ & $C2/m$ & -13.4 \\
                        & $3Q_M$ & $P321$ & -15.5 \\
\midrule
Pure $\Gamma_7^3$ ($L$) & $1Q_L$ & $C2/c$ & -10.7 \\
                        & $2Q_L$ & $C2/m$ & -14.9 \\
                        & $3Q_L$ & $P\bar{3}c1$ & -17.2 \\
\midrule
Mixed ($L+M$)           & $1Q_M+1Q_L$ & $P\bar{1}$ & -14.5 \\
                        & $2Q_M+1Q_L$ & $C2/c$ & -17.7 \\
                        & \textbf{2Q\textsubscript{L} + 1Q\textsubscript{M}} & \textbf{C2} & \textbf{-18.8} \\
\bottomrule
\end{tabular*}
\end{table}

\begin{figure}[tbp]
\centering{\includegraphics[width=\columnwidth]{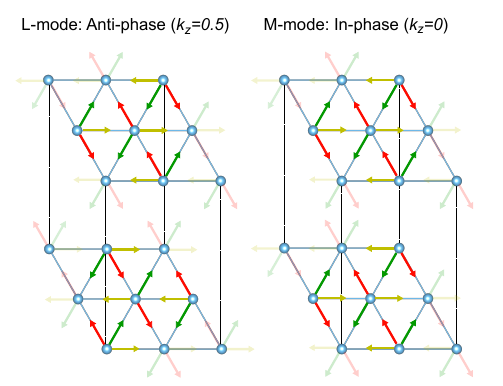}}
\caption{\textbf{Phonon eigenvectors and stacking configurations of the primary lattice instabilities.} \textbf{(Left)} The $L$-point transverse optical phonon mode, characterized by anti-phase interlayer stacking ($k_z=0.5$), drives Ti-trimer formation within each layer through transverse in-plane displacements. \textbf{(Right)} The $M$-point mode exhibits in-phase stacking ($k_z=0$) and describes uniform interlayer shear. The color-coded arrows indicate displacement directions for the three propagation vectors ($\mathbf{q}_1$, $\mathbf{q}_2$, $\mathbf{q}_3$) of the $2\times2\times2$ supercell. The directional selection and cooperative mixing of these modes ($2Q_L + 1Q_M$) stabilize the chiral $C2$ ground state. Se atoms are omitted for clarity.\label{fig:phonon}}
\end{figure}

\begin{figure*}[tbp]
\centering{\includegraphics[width=\textwidth]{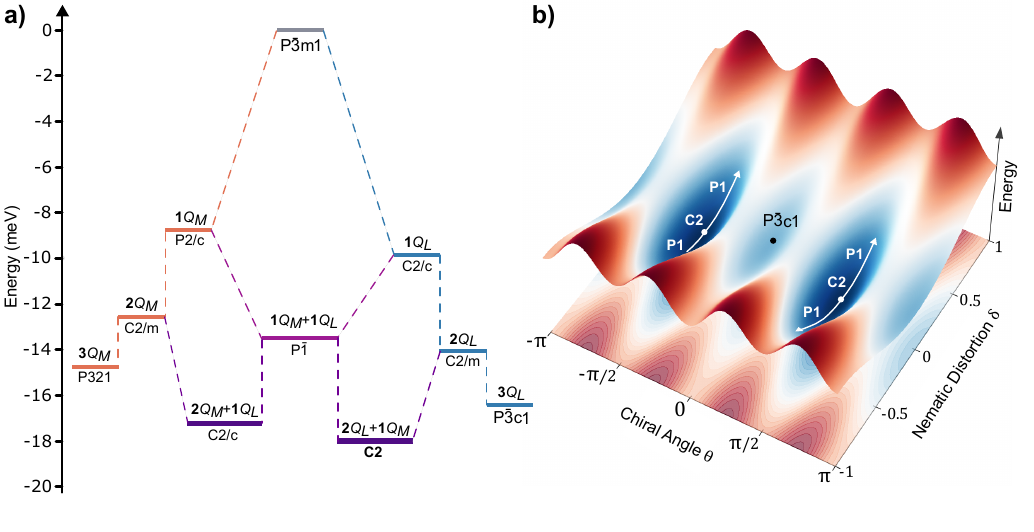}}
\caption{\textbf{\textit{Ab initio} energetic hierarchy and the chiral-nematic energy surface.}
\textbf{(a)} Condensation energy diagram for structural phases derived from $M$-point (orange) and $L$-point (blue) lattice instabilities, relative to the $P\bar{3}m1$ parent phase. The cooperative $2Q_L + 1Q_M$ condensation (purple) establishes the chiral $C2$ configuration as the global energetic minimum, separated from the canonical achiral $P\bar{3}c1$ ($3Q_L$) state by an activation barrier.
\textbf{(b)} Potential energy landscape parameterized by the chiral mixing angle ($\theta$) and the transverse nematic distortion ($\delta$). The topology exhibits a double-well structure along $\theta$, separating left- and right-handed $C2$ ground states ($\theta = \pm\pi/2$) by the $P\bar{3}c1$ saddle point ($\theta = 0$). Along the $\delta$-axis, the landscape exhibits a flat soft manifold (white arrows), indicating the accessibility of triclinic $P1$ configurations at low energetic cost.\label{fig:energetic_landscape}}
\end{figure*}

To map the energy landscape of possible lattice distortions, we start from the primary lattice instabilities of the parent phase, that is, the $L$-point and $M$-point transverse optical modes hybridizing with the acoustic branches, which were identified in previous phonon studies~\cite{Calandra2011, Subedi2022, web11}. The $L$-mode corresponds to transverse in-plane displacements driving Ti-trimer formation with anti-phase interlayer stacking ($k_z=0.5$), while the $M$-mode corresponds to in-phase stacking ($k_z=0$) (Figure~\ref{fig:phonon}). We construct candidate structures by combining these $L$ and $M$ distortions across one, two, or all three propagation directions ($\mathbf{q}_1, \mathbf{q}_2, \mathbf{q}_3$). This generates a complete hierarchy of candidate phases spanning various symmetries, as detailed in Table~\ref{tab:energetics}.

\begin{figure}[tbp]
\centering{\includegraphics[width=\columnwidth]{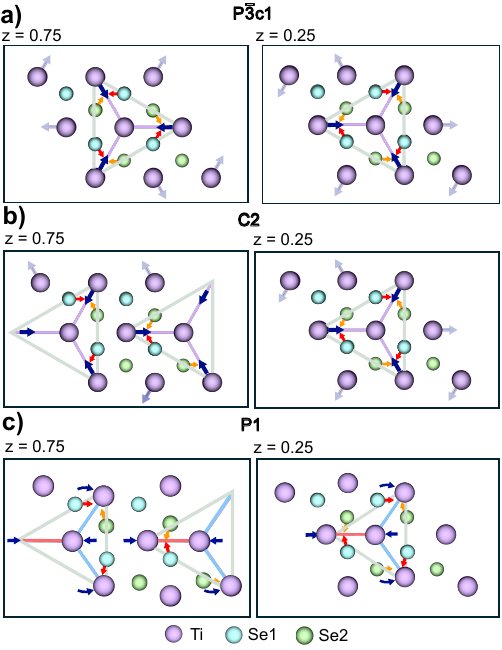}}
\caption{\textbf{Real-space structural decomposition of the charge density wave models.} Top-down view of the 2D projected atomic displacements for Ti layers at $z = 0.75$ and $z = 0.25$. Ti atoms are shown in purple, Se atoms in the top layer (Se1) in blue, and Se atoms in the bottom layer (Se2) in green.
\textbf{(a)} The canonical achiral $P\bar{3}c1$ state driven by an isotropic $3Q_L$ condensation, featuring symmetric Ti-trimer formations that preserve three-fold rotational and inversion symmetries.
\textbf{(b)} The chiral $C2$ ground state driven by the $2Q_L + 1Q_M$ directional selection. The superposition of anti-phase trimerization ($L$-mode) and uniform interlayer shear ($M$-mode) establishes a unique crystallographic axis and co-aligns the Ti-trimers, explicitly breaking both three-fold rotational and inversion symmetries.
\textbf{(c)} The triclinic $P1$ phase resulting from nematic fluctuations along the soft $\delta$-coordinate. Amplitude imbalance between the two active $L$-modes induces intra-trimer bond disproportionation, breaking the remaining two-fold rotation axis, as highlighted by the contracted (red) and stretched (blue) Ti-Ti bonds.\label{fig:structures}}
\end{figure}

To understand how these modes compete and mix, we introduce a Landau free-energy expansion. 
The total free energy $F = F_L + F_M + F_{int}$ is constructed from energetic contributions from individual modes and the interactions between them. The leading interaction ($F_{int}$) between the $L$- and $M$-mode amplitudes on the $i$-th propagation axis ($Q_{L,i}$ and $Q_{M,i}$) is governed by fourth-order interaction terms:
\begin{equation}\label{eq:F_int}
F_{\text{int}} = \kappa \sum_{i} Q_{L,i}^2 Q_{M,i}^2 + \gamma \sum_{i \neq j} Q_{L,i}^2 Q_{M,j}^2 .
\end{equation}
Fitting our \textit{ab initio} total energies to this expansion yields $\kappa > \gamma > 0$. 
The large positive $\kappa$ represents a severe energetic penalty for simultaneous condensation of both $L$ and $M$-mode distortions on the same axis. 
This is simply because the anti-phase ($L$-mode) and in-phase ($M$-mode) trimerizations compete for the same in-plane coordinate. 
To minimize the penalty while maximizing the condensation energy of both modes, the lattice strictly avoids mixed modes on any single axis. 
The term $\gamma$ governs the weaker cross-axis coupling.
Because our calculated coefficients show the system energetically favors $L$-mode condensation over the $M$-mode (see Supplemental Material, Sec.~\ref{secS3}), it selects an asymmetric configuration of anti-phase trimerization on two axes and in-phase on the third ($2Q_L+1Q_M$). 
This specific configuration eliminates the $\kappa$ penalty entirely and co-aligns the Ti-trimers along a single crystallographic axis. 
By breaking both three-fold rotational and inversion symmetries, this cooperative condensation naturally establishes the chiral $C2$ structure as the lowest-energy state (Figure~\ref{fig:structures}b and Table~\ref{tab:energetics}). 
Although another low energy structure (Table~\ref{tab:energetics} and Figure~\ref{fig:energetic_landscape}a), $C2/c$, exists with a proximal energy difference, a transition from $C2$ to $C2/c$ is highly unfavorable: it would require completely suppressing the distortion along one axis before rebuilding it on another. 
The corresponding energy barrier is equivalent to the energy needed to melt the distortion along an entire direction, preventing such a transition and trapping the system in the $C2$ energy-minimum structure upon cooling.

The relative sign of the $2Q_L$ and $1Q_M$ amplitudes dictates the specific chirality (right- vs. left-handedness) of the $C2$ structures. To see how these structures are connected, we map the energy landscape using two collective coordinates (Figure~\ref{fig:energetic_landscape}b). The first is the chiral mixing angle ($\theta$). It describes the distortion path that flips the system's handedness.
Interestingly, the shortest path between these $C2$ states passes through the achiral $P\bar{3}c1$ structure, which serves as the saddle point. 
The second is the nematic distortion ($\delta$), defined mathematically as the amplitude imbalance between the two active transverse $L$-modes. Physically, this distortion manifests as the unequal stretching and compressing of bonds within the Ti-trimers. This bond-stretching distortion breaks the remaining two-fold rotational symmetry of the monoclinic lattice, resulting in a fully asymmetric, triclinic $P1$ configuration.
This coordinate space describes only continuous local distortions within the $2Q_L+1Q_M$ framework; reaching the other phases in Table~\ref{tab:energetics} would require discontinuously switching the modes to different propagation axes.

Crucially, our calculations reveal that the energy landscape is extremely flat along the bond-stretching $\delta$ coordinate, yet with a single minimum at $C2$ (Figure~\ref{fig:energetic_landscape}b). Displacements along $\delta$ cost an energy of 1~meV or less per supercell over the range explored by the fluctuations (Table~\ref{tabS8} in the Supplemental Material), far below the thermal energy at the transition, and a configuration displaced as far as $\delta = 0.31$ relaxes back to $C2$. The lattice can therefore fluctuate freely into $P1$ arrangements around the chiral minimum, while leaving the higher barrier separating the two handednesses ($\theta$) untouched. The $P1$ structure is thus not a competing phase but the symmetry of instantaneous structures visited by the fluctuating chiral lattice. Quantizing $\delta$ as a single coordinate in this potential (Sec.~\ref{secS4} of the Supplemental Material) yields a ground state that is exactly symmetric, $\langle\delta\rangle = 0$, yet spreads broadly across the soft well through zero-point motion alone.

This picture reconciles the seemingly contradictory symmetries reported in experiments. The static $C2$ lattice has already lost inversion and three-fold rotational symmetries, which qualitatively accounts for the principal spectroscopic signatures, such as the splitting of degenerate optical phonon modes and their infrared activation. The fluctuations additionally remove the remaining twofold axis; because the spectroscopic fingerprints of this distortion grow with $|\delta|$ (mode splittings) or $\delta^2$ (mode intensities), they persist in the measured averages even though the average distortion vanishes, and vibrational probes can therefore detect no remaining point symmetry. Probes that instead average the atomic positions directly---diffraction within a uniform chiral domain, and scanning tunneling microscopy, whose topographic contrast is insensitive to $\delta$ (Figure~\ref{fig:sts}a)---recover the chiral $C2$ structure.

Having established this lattice-driven $C2$ ground state, we can address two apparent tensions with prior literature. First, recent models propose that chirality originates from ``charge-lattice frustration,'' invoking an independent electronic charge modulation, $\delta \rho_L$, that is symmetry-mismatched with the lattice~\cite{Kim2024}. Standard DFT structural relaxations, however, couple the electronic density and ionic positions self-consistently, not allowing an electronic order parameter that is variationally free from the lattice. Within this self-consistent framework, we examined whether any lattice proxy for the proposed electronic modulation appears in the relaxed structure. Decomposing the $C2$ ground state into its symmetry components reveals that the Se-only displacement associated with this mode is nearly two orders of magnitude smaller than the primary Ti-centered lattice distortions (Table~\ref{tabS9} in the Supplemental Material). This indicates that, within the standard adiabatic framework, the $2Q_L+1Q_M$ mechanism provides a sufficient lattice-driven origin without requiring an additional frustration term. Second, inelastic X-ray scattering shows that the $M$-point phonon remains hard at $T_c$~\cite{hol01,web11,Kim2024}. This observation is not incompatible with the $C2$ structure, because the relaxation proceeds dynamically within the single phase transition. The dominant $L$-mode condensation initially forms a $P\bar{3}c1$ configuration that is itself unstable to transverse shear distortions. The lattice slides toward the $C2$ global minimum along this shear coordinate without requiring the $M$ phonon to independently soften or to drive a separate thermodynamic transition.
The participation of the $M$-point phonon in the transition through the shear mode is fully consistent with the previous experimental works~\cite{web11,Kim2024}.

\subsection{Comparison with experimental observations}

\begin{figure*}[tbp]
\centering{\includegraphics[width=\textwidth]{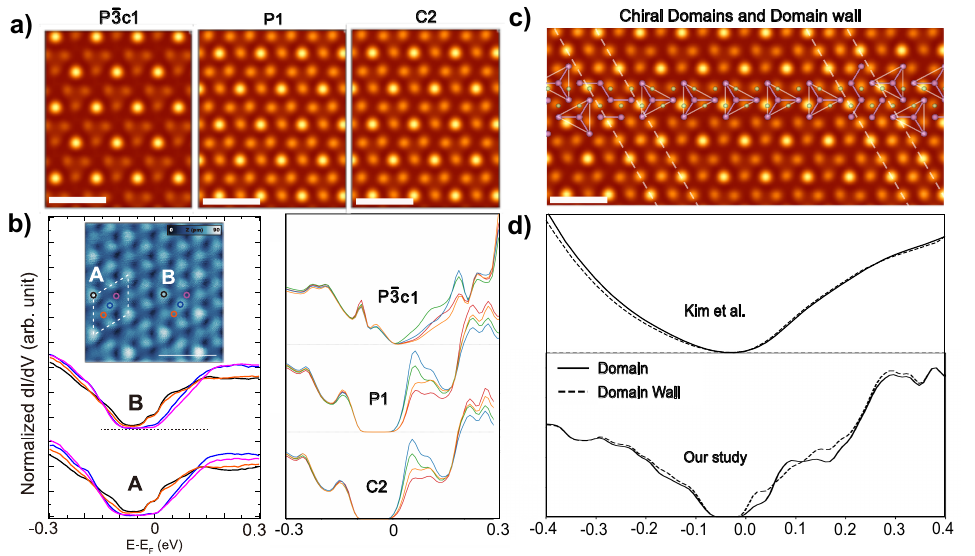}}
\caption{\textbf{Simulated and experimental STM/STS.}
\textbf{(a)} Simulated constant-current STM topographies ($+120$ meV). The achiral $P\bar{3}c1$ phase (left) exhibits a hexagonal superlattice, whereas the chiral $C2$ and $P1$ phases display nearly indistinguishable triangular motifs. This indistinguishable STM profile reflects the subtle intra-trimer bond disproportionation of the nematic fluctuation within the $C2$ framework.
\textbf{(b)} Experimental (left) and calculated (right) local electronic structure. Left: STM image ($-0.2$ V, $200$ pA, $2.4\times 2.4$ nm$^2$) with spatially resolved dI/dV spectra (positions A and B). For dI/dV spectra measured on location B, vertical offsets are given for comparison (dashed line). Right: LDOS for the three candidate phases. The chiral $C2$ and $P1$ phases open a hard gap ($\sim$80--100 meV), while $P\bar{3}c1$ remains pseudogapped near $E_F$.
\textbf{(c)} Simulated topography with atomic overlay of a seamless domain wall connecting antipodal chiral $C2$ domains (white dashed lines). The half-unit-cell translational shift preserves trimerization and stacking topology across the boundary.
\textbf{(d)} LDOS across the domain wall. Top: Experimental STS data adapted from Ref.~\cite{kim2024nanolett} showing preserved gap magnitude in the domain (solid) and at the wall (dashed). Bottom: Calculated LDOS for the $C2$ wall, reproducing the absence of mid-gap states despite local $1Q$ symmetry.\label{fig:sts}}
\end{figure*}

Although total energy calculations show that the chiral $C2$ structure is stable, this model must also reproduce the experimental electronic and spatial signatures without relying on independent electronic orders. Previous work established that the $C2$ structure excellently reproduces the chiral topography of STM images~\cite{kim2024nanolett}. Here, we also simulated the STM topography for the triclinic $P1$ configurations. Since the $P1$ symmetry represents a continuous path along this soft manifold rather than a single fixed minimum, we selected a representative configuration ($\delta \approx 0.3$, Table~\ref{tabS6} in the Supplemental Material). As shown in Figure~\ref{fig:sts}a, this highly asymmetric $P1$ configuration reproduces the chiral STM topography just as well as the $C2$ ground state. The nearly indistinguishable topographies show that the STM contrast is generated by the structural motif itself and is insensitive to $\delta$: as a surface-sensitive probe, STM images the chiral order irrespective of the instantaneous configuration of the fluctuating lattice, whereas vibrational spectroscopies respond to the instantaneous local bonding environment. The two apparently exclusive symmetry assignments therefore describe different averages of the same fluctuating lattice, rather than two different phases. 

We next examine the single-particle electronic structure of the competing phases. The achiral $P\bar{3}c1$ phase exhibits a semimetallic local density of states (LDOS), which is inconsistent with our spatially resolved dI/dV spectra (Figure~\ref{fig:sts}b) and previous measurements~\cite{kim2024nanolett}. In sharp contrast, the chiral $C2$ distortion spontaneously generates an insulating surface gap ($\sim$100 meV). Previous theoretical models required empirical electronic correlation corrections to reproduce this gap magnitude~\cite{kim2024nanolett}. Our calculations demonstrate that a hard gap emerges from first-principles without such \textit{ad hoc} parameters. This indicates that the lattice symmetry breaking provides the primary structural driving force for the insulating state, even if correlation effects are ultimately required for a quantitative match (details on functional benchmarking are provided in the Supplemental Material, Sec.~\ref{secS1}). Consistent with this picture, self-consistent many-body calculations including excitonic correlations find an exciton binding energy an order of magnitude smaller than the gap, and show that insulating spectra arise from lattice distortions, static or fluctuating~\cite{Pashov2025}; likewise, the pseudogap and replica-band signatures of the normal phase are reproduced by dynamical electron-phonon self-energies without excitonic terms~\cite{Fragkos2026Arxiv}. Furthermore, the calculated bulk band structure yields a smaller indirect gap ($\sim$20--30 meV, Fig.~\ref{fig:arpes} in the Supplemental Material), which is consistent with bulk angle-resolved photoemission spectroscopy measurements~\cite{Watson2019,ghafari2017}. Band dispersions sampled along the nematic coordinate ($\delta \lesssim 0.3$) preserve this insulating topology, indicating that the dynamic structural fluctuations do not alter the fundamental electronic character.

The behavior of the $C2$ ground state can be further examined at topological defects.
Figure~\ref{fig:sts}c illustrates a domain wall that separates connecting antipodal (opposite-handed) chiral domains. At this boundary, the Ti-trimer network is locally disrupted, forming a $1Q_L$-like linear structure to bridge the two domains. Despite this structural disruption, our calculated local density of states (LDOS) indicates that the energy gap remains open across the boundary (Figure~\ref{fig:sts}d). This preserved gap is consistent with the experimental tunneling spectroscopy reported by Kim \textit{et al.}~\cite{kim2024nanolett}, which shows an absence of mid-gap states at the wall. This insulating nature distinguishes 1$T$-TiSe$_2$ from the domain walls with distinct in-gap states as reported in conventional Peierls systems like 1$T$-TaS$_2$~\cite{Park2021}. Furthermore, our modeling suggests that this gapped boundary behavior remains stable in the presence of nematic fluctuations ($\delta \neq 0$), as the intra-trimer bond stretching along the soft manifold does not alter the underlying geometric constraints of the boundary orientations.

\section{Discussion}\label{sec3}

These findings suggest that the microscopic origin of the insulating state in 1$T$-TiSe$_2$ is consistent with a lattice-driven mechanism.
By identifying the chiral $C2$ phase as the global minimum within a soft nematic manifold, we demonstrate that cooperative condensation of primary $L$- and $M$-point phonon modes generates sufficient band reconstruction to open an insulating gap consistent with experiments.
While quantitative accuracy in the gap magnitude requires the meta-GGA $r^2\text{SCAN}+r\text{VV10}$ functional, the qualitative trend is robust across different approximations. Specifically, the chiral $C2$ phase consistently opens a hard gap while the achiral $P\bar{3}c1$ phase remains metallic in standard semilocal GGAs (PBE+D3, optB86b; Fig.~\ref{fig:benchmark} in the Supplemental Material).
This functional independence indicates that the existence of the CDW gap is primarily controlled by the chiral $C2$ symmetry breaking, while nonlocal exchange and excitonic correlations likely control its precise magnitude and fine structure~\cite{hellgren2017critical,bianco2015electronic,chen2015charge}.

The soft nematic energy surface unifies the divergent experimental pictures of the TiSe$_2$ ground state.
The flat potential along the nematic coordinate $\delta$ allows the dynamical fluctuation into the triclinic $P1$ configuration at a sub-meV energetic cost.
A related mechanism operates in the monolayer limit, where long-wavelength acoustic-phonon fluctuations on an asymmetric potential surface generate unequal amplitudes of the three charge-density-wave components, stabilizing a chiral average structure at low temperature, while relaxing the fluctuation-averaged structure recovers the symmetric achiral form~\cite{Benic2026}. In the bulk phase studied here, the directional $2Q_L + 1Q_M$ mixing is instead selected statically at $T = 0$ by the quartic mode coupling, and the fluctuations subsequently explore the soft $\delta$ direction around the chiral minimum. The two mechanisms are complementary: both act on near-flat, asymmetric potential energy surfaces, and both yield average structures whose symmetry is higher than that of the instantaneous configurations.
While the chiral $C2$ phase is established as the definitive $T=0$ ground state, the shallow energy landscape and sub-meV nematic fluctuations suggest that finite-temperature properties may be sensitive to lattice anharmonicity.
The temperature-dependent phonon renormalization, including the evolution of the $M$-mode response, falls outside the scope of the present static zero-temperature framework.
\textit{Ab initio} molecular dynamics or finite-temperature lattice dynamics would be valuable extensions to explore whether anharmonic stabilization effects contribute to these experimental signatures.

The small energy difference between the chiral $C2$ global minimum and the achiral $P\bar{3}c1$ saddle point ($\Delta E \approx 1.6$ meV/supercell) in the present free energy surface can consistently interpret very recent small-area-electron-diffraction measurements~\cite{Wang2026}. 
This work reported bulk phase separation into regions with and without $M$-point diffraction peaks.
Because the energetic cost to flip the chirality is so small, the pristine bulk is highly susceptible to nanoscale domain fragmentation upon cooling. 
If the bulk forms a dense mosaic of static $C2$ domains that are smaller than the electron beam's transverse coherence length, the out-of-phase $+1Q_M$ and $-1Q_M$ components would destructively interfere, effectively extinguishing the elastic $M$-point diffraction peaks. 
Conversely, local pinning due to defects or strain gradients would stabilize macroscopic $C2$ domains larger than the coherence length, allowing the $M$-peaks to emerge. 
Within a uniform chiral domain, coherently diffracting probes average to the $C2$-centered structure; the achiral average structure resolved by early crystallographic analyses then follows naturally if the sample contains comparable populations of left- and right-handed domains, which is also a statistical picture independently proposed on theoretical grounds~\cite{Carpene2026}.
This may explain why surface-sensitive probes like STM capture a chiral configuration, possibly unifying apparently contradictory surface and bulk measurements.

\section{Summary}
Chiral charge-density waves in 1$T$-TiSe$_2$ can be understood as a lattice-driven instability in which cooperative condensation of $L$- and $M$-point phonon modes stabilizes a monoclinic $C2$ ground state and opens an insulating gap consistent with STM and ARPES experiments. This $C2$ phase sits at the center of a soft nematic energy manifold, a single shallow valley whose quantum ground state is exactly centered ($\langle\delta\rangle = 0$) yet broad enough to visit fully asymmetric $P1$ arrangements even at zero temperature. A chiral lattice fluctuating over this shallow well reconciles the two experimental pictures that have long appeared contradictory: structure probes (such as STM topography and diffraction within a uniform chiral domain) recover the average $C2$ lattice, while vibrational spectroscopies measure the instantaneous local symmetry and find none remaining. More broadly, this mechanism shows how chiral-nematic CDWs can emerge from the interplay of chiral lattice instabilities and nematic fluctuations, providing a concrete structural framework for interpreting symmetry breaking in TiSe$_2$ and related quantum materials.

\section{Methods}

\textbf{Computational Details.} Density functional theory calculations were performed using the projector augmented-wave (PAW) method~\cite{kresse1999} as implemented in the \textsc{VASP} code~\cite{kresse1996, kresse199615}. Exchange-correlation interactions were treated using the $r^2$SCAN+rVV10 functional~\cite{Furness2020, Sabatini2013, Ning2022} with a 520~eV plane-wave cutoff. We used $\Gamma$-centered $k$-point meshes of $12\times12\times6$ for the primitive cell, with correspondingly reduced grids for the $2\times2\times2$ supercells. Structural optimizations were performed with convergence criteria of $10^{-8}$~eV for total energy and $10^{-3}$~eV/Å for atomic forces. Spin-orbit coupling was included in electronic structure calculations for accurate determination of band gaps and dispersions, while structural energetics were found to be negligibly affected by relativistic effects. To test the lattice instability mechanism, we generated initial atomic distortions corresponding to the $L$-point ($\Gamma_7^3$) and $M$-point ($\Gamma_5^3$) irreducible representations identified in Kim et al.~\cite{Kim2024}, following the instability analysis by Subedi~\cite{Subedi2022}. Symmetry identification was performed using the \textsc{pymatgen} library~\cite{Ong2013}. Scanning tunneling microscopy images were simulated within the Tersoff-Hamann approximation~\cite{Tersoff1998}. Data were processed using \textsc{VASPKIT}~\cite{Wang2021}.

\textbf{Experimental Details.} Atomically resolved imaging and spectroscopy were performed under ultrahigh vacuum (UHV) at 2.2 K using a commercial low-temperature scanning tunneling microscope (STM) and Pt-Ir tips. For scanning tunneling spectroscopy (STS) measurements, the dI/dV spectra were recorded with the lock-in technique using a 15 mV modulation voltage. Clean 1$T$-TiSe$_2$ surface was prepared by UHV cleavage at room temperature.


\begin{acknowledgments}
This work was supported by the Institute for Basic Science (Grant No. IBS-R014-D1).
H.W.Y. conceived the project, directed the overall narrative of the research, and edited and finalized the manuscript. A.A.A. performed the \textit{ab initio} calculations, derived the theoretical model, wrote the original draft of the manuscript, and prepared the figures. E.D. conducted the STM/STS experiments and provided all experimental data. J.W.P. provided project supervision and scientific guidance, and reviewed the manuscript.
The authors declare no conflicts of interest.
\end{acknowledgments}

\section*{Data Availability}
The data that support the findings of this study are available from the corresponding author upon reasonable request.

\bibliography{references}

\clearpage
\onecolumngrid
\setcounter{section}{0}
\setcounter{figure}{0}
\setcounter{table}{0}
\setcounter{equation}{0}
\renewcommand{\thesection}{S\arabic{section}}
\renewcommand{\thesubsection}{\thesection.\arabic{subsection}}
\renewcommand{\thefigure}{S\arabic{figure}}
\renewcommand{\thetable}{S\arabic{table}}
\renewcommand{\theequation}{S\arabic{equation}}

\begin{center}
  {\large\bfseries Supplemental Material for\\[0.3em]
  ``Chiral-Nematic Lattice Instability in Charge Density Wave of 1$T$-TiSe$_2$''\par}
  \vspace{0.6em}
  {\normalsize A.~A.~Alghifari, E.~Do, J.~W.~Park, and H.~W.~Yeom\par}
  \vspace{0.6em}
\end{center}
\noindent References cited in this Supplemental Material are listed in the bibliography of the main text.

\section{Functional Independence of the Structural Gap}\label{secS1}

\begin{figure}[htbp]
\centering
\includegraphics[width=0.7\textwidth]{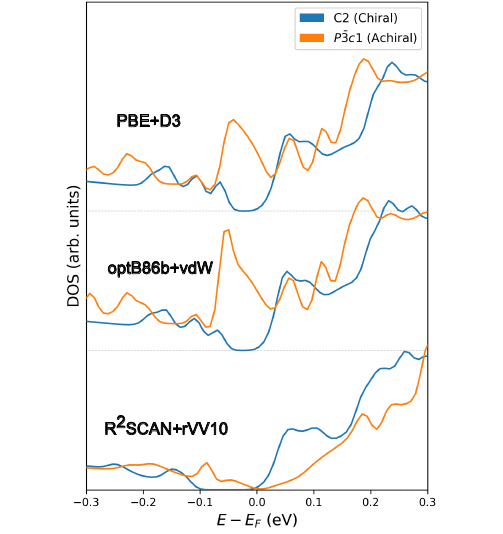}
\caption{\textbf{Integrated Density of States for different functionals.} PBE+D3 and optB86b+vdW show a similar curve to each other but different to $r^2$SCAN+rVV10. Despite being metallic in achiral phase ($P\bar{3}c1$) compared to semimetallic in $r^2$SCAN+rVV10, all tested functionals show a hard gap opening in the chiral phases ($C2$ and $P1$), with $\sim$30 meV and $\sim$80--100 meV for GGA and meta-GGA, respectively. The selected GGA functionals also exhibit better qualitative agreement with meta-GGA in the chiral phases compared to the achiral one.\label{fig:benchmark}}
\end{figure}

Isolating the structural contribution to the insulating gap requires demonstrating that the gap opens in the chiral phase even in the absence of non-local exchange or empirical correlations. Figure~\ref{fig:benchmark} compares the electronic density of states (DOS) for the achiral $P\bar{3}c1$ (orange) and chiral $C2$ (blue) phases across three functionals: semilocal PBE+D3~\cite{perdew1996generalized, grimme2010consistent} and optB86b-vdW~\cite{klimes2011van}, and the meta-GGA $r^2$SCAN+rVV10~\cite{Furness2020, peng2016r2scan}.

For standard GGAs (PBE+D3, optB86b-vdW), the $P\bar{3}c1$ phase exhibits finite spectral weight at the Fermi level ($E_F$), consistent with metallic character, whereas the $C2$ phase opens a distinct gap of approximately 30 meV. The $r^2$SCAN+rVV10 functional predicts a semimetallic $P\bar{3}c1$ phase with suppressed DOS at $E_F$, while opening a larger hard gap of $\sim$80--100 meV in the $C2$ phase. The emergence of a gap exclusively in the chiral phase across all functionals---despite their inability to capture excitonic correlations---confirms that the lattice symmetry breaking alone is sufficient to drive the insulating transition.

Table~\ref{tabS1} provides the corresponding lattice parameter benchmarks.

\begin{table}[htbp]
\centering
\caption{Comparison of Lattice parameters and $c$-axis errors for the $\text{TiSe}_2$ normal phase. Data for the TASK+D2, and MVS+D2 functionals are taken from Yin \textit{et al.}~\cite{yin2024}, while HSE06 values are from Hellgren \textit{et al.}~\cite{hellgren2017critical}.\label{tabS1}}
\begin{tabular}{lcccc}
\toprule
Functional & $a$ (\AA) & $c$ (\AA) & $c$-error (\%) \\
\midrule
Experiment & 3.540 & 6.007 & --- \\
PBE+D3 & 3.512 & 5.946 & -1.0 \\
$r^2$SCAN+rVV10 & 3.539 & 6.033 & 0.4 \\
TASK+D2 & 3.659 & 7.456 & 24.2 \\
MVS+D2 & 3.607 & 9.891 & 64.8 \\
HSE06 & 3.540 & 6.104 & 1.6 \\
\bottomrule
\end{tabular}
\end{table}

A parallel functional-independence test can be carried out for the structural energetics. All competing phases of Table~\ref{tab:energetics} of the main text were fully relaxed with the semilocal PBE(+D3) functional~\cite{perdew1996generalized, grimme2010consistent}, using otherwise identical computational settings (Table~\ref{tabS2}). Although the condensation energies are compressed to about two-thirds of the $r^2$SCAN+rVV10 values, the key hierarchy is preserved: at equal propagation-vector count the $L$-mode condensates remain more strongly bound than the $M$-mode ones, and the chiral $C2$ phase remains the global minimum, lying $\sim$0.5~meV below $P\bar{3}c1$ per supercell (versus 1.6~meV with $r^2$SCAN+rVV10).

\begin{table}[htbp]
\centering
\caption{Phase energetics for two exchange--correlation functionals,
relative to the pristine $P\bar{3}m1$ structure (meV per $2\times2\times2$
cell). The $r^2$SCAN+rVV10 column reproduces Table~\ref{tab:energetics} of the main
manuscript for direct comparison.\label{tabS2}}
\begin{tabular}{llccc}
\toprule
Family & Distortion & Symmetry & $\Delta E$ ($r^2$SCAN+rVV10) & $\Delta E$ (PBE+D3) \\
\midrule
Reference & --- & $P\bar{3}m1$ & 0.00 & 0.00 \\
Pure $M$ & $1Q_M$ & $P2/c$ & $-9.7$ & $-7.8$ \\
         & $2Q_M$ & $C2/m$ & $-13.4$ & $-8.1$ \\
         & $3Q_M$ & $P321$ & $-15.5$ & $-8.3$ \\
Pure $L$ & $1Q_L$ & $C2/c$ & $-10.7$ & $-8.7$ \\
         & $2Q_L$ & $C2/m$ & $-14.9$ & $-10.7$ \\
         & $3Q_L$ & $P\bar{3}c1$ & $-17.2$ & $-12.0$ \\
Mixed    & $1Q_M+1Q_L$ & $P\bar{1}$ & $-14.5$ & $-8.4$ \\
         & $2Q_M+1Q_L$ & $C2/c$ & $-17.7$ & $-10.7$ \\
         & $2Q_L+1Q_M$ & $C2$ & $-18.8$ & $\mathbf{-12.5}$ \\
\bottomrule
\end{tabular}
\end{table}

\section{Ground-State and Metastable Crystal Structures}

We provide fully relaxed structural parameters for the phases analyzed in the energetic hierarchy. All coordinates are given in fractional units of the lattice vectors. Wyckoff positions represent the approximate site symmetry corresponding to the ideal high-symmetry prototype; coordinates deviate slightly from ideal positions due to numerical convergence and the relaxed distortion amplitudes.

\begin{table}[htbp]
\centering
\caption{Structural parameters for the parent phase $P\bar{3}m1$ (\#164). Lattice constants: $a=b=7.07944$~\AA, $c=12.06332$~\AA. Wyckoff positions represent the approximate site symmetry. Coordinates are fully relaxed and deviate from ideal positions due to the reconstruction in the 2$\times$2$\times$2 supercell.\label{tabS3}}
\begin{tabular}{lcccc}
\toprule
Atom & Wyckoff & $x$ & $y$ & $z$ \\
\midrule
Ti(1) & $1a$ & 0.00000 & 0.00000 & 0.25000 \\
Ti(2) & $2d$ & 0.50000 & 0.00000 & 0.25000 \\
Ti(3) & $2d$ & 0.00000 & 0.50000 & 0.25000 \\
Ti(4) & $2d$ & 0.50000 & 0.50000 & 0.25000 \\
Ti(5) & $1b$ & 0.00000 & 0.00000 & 0.75000 \\
Ti(6) & $2d$ & 0.50000 & 0.00000 & 0.75000 \\
Ti(7) & $2d$ & 0.00000 & 0.50000 & 0.75000 \\
Ti(8) & $2d$ & 0.50000 & 0.50000 & 0.75000 \\
Se(1) & $2d$ & 0.16667 & 0.33333 & 0.37606 \\
Se(2) & $2d$ & 0.66667 & 0.33333 & 0.37606 \\
Se(3) & $2d$ & 0.16667 & 0.83333 & 0.37606 \\
Se(4) & $2d$ & 0.66667 & 0.83333 & 0.37606 \\
Se(5) & $2d$ & 0.33333 & 0.16667 & 0.12394 \\
Se(6) & $2d$ & 0.83333 & 0.16667 & 0.12394 \\
Se(7) & $2d$ & 0.33333 & 0.66667 & 0.12394 \\
Se(8) & $2d$ & 0.83333 & 0.66667 & 0.12394 \\
Se(9) & $2d$ & 0.16667 & 0.33333 & 0.87606 \\
Se(10) & $2d$ & 0.66667 & 0.33333 & 0.87606 \\
Se(11) & $2d$ & 0.16667 & 0.83333 & 0.87606 \\
Se(12) & $2d$ & 0.66667 & 0.83333 & 0.87606 \\
Se(13) & $2d$ & 0.33333 & 0.16667 & 0.62394 \\
Se(14) & $2d$ & 0.83333 & 0.16667 & 0.62394 \\
Se(15) & $2d$ & 0.33333 & 0.66667 & 0.62394 \\
Se(16) & $2d$ & 0.83333 & 0.66667 & 0.62394 \\
\bottomrule
\end{tabular}
\end{table}

\begin{table}[htbp]
\centering
\caption{Structural parameters for the canonical phase $P\bar{3}c1$ (\#165). Lattice constants: $a=b=7.07445$~\AA, $c=12.10094$~\AA.\label{tabS4}}
\begin{tabular}{lcccc}
\toprule
Atom & Wyckoff & $x$ & $y$ & $z$ \\
\midrule
Ti(1) & $2a$ & 0.00000 & 0.00860 & 0.25000 \\
Ti(2) & $6f$ & 0.49140 & 0.99140 & 0.25000 \\
Ti(3) & $2b$ & 0.00000 & 0.50000 & 0.25000 \\
Ti(4) & $6f$ & 0.50860 & 0.50000 & 0.25000 \\
Ti(5) & $2a$ & 0.00000 & 0.99140 & 0.75000 \\
Ti(6) & $6f$ & 0.50860 & 0.00860 & 0.75000 \\
Ti(7) & $2b$ & 0.00000 & 0.50000 & 0.75000 \\
Ti(8) & $6f$ & 0.49140 & 0.50000 & 0.75000 \\
Se(1) & $4d$ & 0.16372 & 0.33064 & 0.37615 \\
Se(2) & $12g$ & 0.66692 & 0.33628 & 0.37615 \\
Se(3) & $12g$ & 0.16936 & 0.83308 & 0.37615 \\
Se(4) & $4d$ & 0.66667 & 0.83333 & 0.37560 \\
Se(5) & $4d$ & 0.33333 & 0.16667 & 0.12440 \\
Se(6) & $12g$ & 0.83628 & 0.16692 & 0.12385 \\
Se(7) & $12g$ & 0.33308 & 0.66936 & 0.12385 \\
Se(8) & $12g$ & 0.83064 & 0.66372 & 0.12385 \\
Se(9) & $12g$ & 0.16936 & 0.33628 & 0.87615 \\
Se(10) & $12g$ & 0.66692 & 0.33064 & 0.87615 \\
Se(11) & $12g$ & 0.16372 & 0.83308 & 0.87615 \\
Se(12) & $4d$ & 0.66667 & 0.83333 & 0.87560 \\
Se(13) & $4d$ & 0.33333 & 0.16667 & 0.62440 \\
Se(14) & $12g$ & 0.83064 & 0.16692 & 0.62385 \\
Se(15) & $12g$ & 0.33308 & 0.66372 & 0.62385 \\
Se(16) & $12g$ & 0.83628 & 0.66936 & 0.62385 \\
\bottomrule
\end{tabular}
\end{table}

\begin{table}[htbp]
\centering
\caption{Structural parameters for the chiral ground state $C2$ (\#5). Lattice constants: $a=7.07553$~\AA, $b=7.07553$~\AA, $c=12.10981$~\AA, $\alpha=90.00^\circ$, $\beta=90.01^\circ$, $\gamma=120.01^\circ$.\label{tabS5}}
\begin{tabular}{lcccc}
\toprule
Atom & Wyckoff & $x$ & $y$ & $z$ \\
\midrule
Ti(1) & $4c$ & 0.00000 & 0.00860 & 0.25000 \\
Ti(2) & $4c$ & 0.49107 & 0.99137 & 0.25002 \\
Ti(3) & $4c$ & 0.00000 & 0.49975 & 0.25000 \\
Ti(4) & $4c$ & 0.50893 & 0.50029 & 0.24998 \\
Ti(5) & $4c$ & 0.00000 & 0.99975 & 0.75000 \\
Ti(6) & $4c$ & 0.50893 & 0.00029 & 0.74998 \\
Ti(7) & $4c$ & 0.00000 & 0.50860 & 0.75000 \\
Ti(8) & $4c$ & 0.49107 & 0.49137 & 0.75002 \\
Se(1) & $4c$ & 0.16352 & 0.33066 & 0.37605 \\
Se(2) & $4c$ & 0.66693 & 0.33628 & 0.37603 \\
Se(3) & $4c$ & 0.16945 & 0.83314 & 0.37600 \\
Se(4) & $4c$ & 0.66664 & 0.83319 & 0.37554 \\
Se(5) & $4c$ & 0.33336 & 0.16655 & 0.12446 \\
Se(6) & $4c$ & 0.83648 & 0.16714 & 0.12395 \\
Se(7) & $4c$ & 0.33307 & 0.66934 & 0.12397 \\
Se(8) & $4c$ & 0.83055 & 0.66369 & 0.12400 \\
Se(9) & $4c$ & 0.16945 & 0.33314 & 0.87600 \\
Se(10) & $4c$ & 0.66664 & 0.33319 & 0.87554 \\
Se(11) & $4c$ & 0.16352 & 0.83066 & 0.87605 \\
Se(12) & $4c$ & 0.66693 & 0.83628 & 0.87603 \\
Se(13) & $4c$ & 0.33307 & 0.16934 & 0.62397 \\
Se(14) & $4c$ & 0.83055 & 0.16369 & 0.62400 \\
Se(15) & $4c$ & 0.33336 & 0.66655 & 0.62446 \\
Se(16) & $4c$ & 0.83648 & 0.66714 & 0.62395 \\
\bottomrule
\end{tabular}
\end{table}

\begin{table}[htbp]
\centering
\caption{Structural parameters for the extreme nematic limit $P1$ (\#1) corresponding to $\delta=0.31$. Lattice constants: $a=7.07557$~\AA, $b=7.07557$~\AA, $c=12.10977$~\AA, $\alpha=90.01^\circ$, $\beta=89.99^\circ$, $\gamma=119.99^\circ$.\label{tabS6}}
\begin{tabular}{lcccc}
\toprule
Atom & Wyckoff & $x$ & $y$ & $z$ \\
\midrule
Ti(1) & $1a$ & 0.99776 & 0.00692 & 0.25001 \\
Ti(2) & $1a$ & 0.49284 & 0.99308 & 0.25001 \\
Ti(3) & $1a$ & 0.00220 & 0.50198 & 0.25000 \\
Ti(4) & $1a$ & 0.50720 & 0.49802 & 0.24998 \\
Ti(5) & $1a$ & 0.00220 & 0.00198 & 0.75000 \\
Ti(6) & $1a$ & 0.50720 & 0.99802 & 0.74998 \\
Ti(7) & $1a$ & 0.99776 & 0.50692 & 0.75001 \\
Ti(8) & $1a$ & 0.49284 & 0.49308 & 0.75001 \\
Se(1) & $1a$ & 0.16419 & 0.33120 & 0.37590 \\
Se(2) & $1a$ & 0.66614 & 0.33558 & 0.37602 \\
Se(3) & $1a$ & 0.16903 & 0.83243 & 0.37603 \\
Se(4) & $1a$ & 0.66732 & 0.83392 & 0.37562 \\
Se(5) & $1a$ & 0.33408 & 0.16725 & 0.12438 \\
Se(6) & $1a$ & 0.83576 & 0.16632 & 0.12397 \\
Se(7) & $1a$ & 0.33236 & 0.66895 & 0.12395 \\
Se(8) & $1a$ & 0.83112 & 0.66435 & 0.12413 \\
Se(9) & $1a$ & 0.16903 & 0.33243 & 0.87603 \\
Se(10) & $1a$ & 0.66732 & 0.33392 & 0.87562 \\
Se(11) & $1a$ & 0.16419 & 0.83120 & 0.87590 \\
Se(12) & $1a$ & 0.66614 & 0.83558 & 0.87602 \\
Se(13) & $1a$ & 0.33236 & 0.16895 & 0.62395 \\
Se(14) & $1a$ & 0.83112 & 0.16435 & 0.62413 \\
Se(15) & $1a$ & 0.33408 & 0.66725 & 0.62438 \\
Se(16) & $1a$ & 0.83576 & 0.66632 & 0.62397 \\
\bottomrule
\end{tabular}
\end{table}
\subsection{Quantitative Structural Metrics}
Table~\ref{tabS7} summarizes the bond length distribution within the Ti-sublattice for the analyzed phases. The parent phase $P\bar{3}m1$ exhibits uniform Ti-Ti distances (3.540 \AA) corresponding to the high-symmetry hexagonal lattice. The distorted phases exhibit significant bond contraction. The $P\bar{3}c1$ and $C2$ phases show uniform short bonds centered at 3.476 \AA, representing a $\sim$1.8\% contraction relative to the parent phase that drives Ti-trimer formation. The nematic $P1$ configuration exhibits bond length disproportionation within the trimer unit ($\Delta d \approx 0.018$ \AA), confirming the structural signature of the soft nematic manifold explored in the main text.

\begin{table}[htbp]
\centering
\caption{Ti-sublattice nearest-neighbor bond length distribution ($d_{\text{Ti-Ti}}$). The parent phase distance is 3.540 \AA. $\Delta d$ indicates the spread of bonds within the trimer unit.\label{tabS7}}
\begin{tabular}{lcccc}
\toprule
Phase & Space Group & $d_{\text{Ti-Ti}}$ Range (\AA) & $\Delta d$ (\AA) & Contraction (\%) \\
\midrule
Parent ($P\bar{3}m1$) & \#164 & 3.540 (Uniform) & 0.000 & 0.0 \\
Canonical ($P\bar{3}c1$) & \#165 & 3.476 (Uniform) & 0.000 & -1.8 \\
Chiral ($C2$) & \#5 & 3.477 (Uniform) & 0.000 & -1.8 \\
Nematic ($P1$) & \#1 & 3.468 -- 3.486 & 0.018 & -1.8 \\
\bottomrule
\end{tabular}
\end{table}

\begin{table}[htbp]
\centering
\caption{Energy cost and structural forces along the nematic distortion coordinate. The dimensionless nematic parameter $\delta = (d_{L,2} - d_{L,3})/(d_{L,2} + d_{L,3})$ quantifies the amplitude imbalance between the active transverse $L$-modes. This theoretical coordinate is mapped to physical dimensions via the intra-trimer bond disproportionation ($\Delta d$) and the maximum absolute atomic displacement ($\Delta R_{max}$) relative to the $C2$ ground state. The maximum force ($F_{max}$) indicates the magnitude of atomic forces prior to tight relaxation; all configurations in this range relax back to the $C2$ minimum without encountering any other minimum.\label{tabS8}}
\small\setlength{\tabcolsep}{4pt}\begin{tabular}{ccccc}
\toprule
Nematic Parameter & Bond Disproportionation & Max Atomic Disp. & Energy Cost & Maximum Force \\
$\delta$ & $\Delta d$ (m\AA) & $\Delta R_{max}$ (m\AA) & $\Delta E$ (meV) & $F_{max}$ (eV/\AA) \\
\midrule
0.3100 & 18.1 & 17.7 & 1.69 & 0.020 \\
0.2625 & 12.5 & 15.6 & 1.04 & 0.018 \\
0.1897 & 10.1 & 12.4 & 0.50 & 0.011 \\
0.1558 & 8.8  & 10.7 & 0.33 & 0.010 \\
0.1221 & 7.2  & 8.8  & 0.21 & 0.009 \\
0.0861 & 5.4  & 6.6  & 0.12 & 0.007 \\
0.0666 & 4.3  & 5.9  & 0.09 & 0.006 \\
0.0457 & 3.2  & 4.6  & 0.05 & 0.005 \\
0.0150 & 2.4  & 2.8  & 0.02 & 0.004 \\
0.0000 & 1.8$^{\text{a}}$ & 0.0  & 0.00 & 0.000 \\
\bottomrule
\end{tabular}
\par
\vspace{1ex}
\small $^{\text{a}}$The residual 1.8 m\AA\ disproportionation in the nominal ground state reflects the finite force-convergence tolerance ($10^{-3}$ eV/\AA) of the DFT structural relaxation, consistent with the slight $L$-mode asymmetry noted in Table~\ref{tabS9}.
\end{table}

\begin{figure}[htbp]
    \centering
    \includegraphics[width=0.7\textwidth]{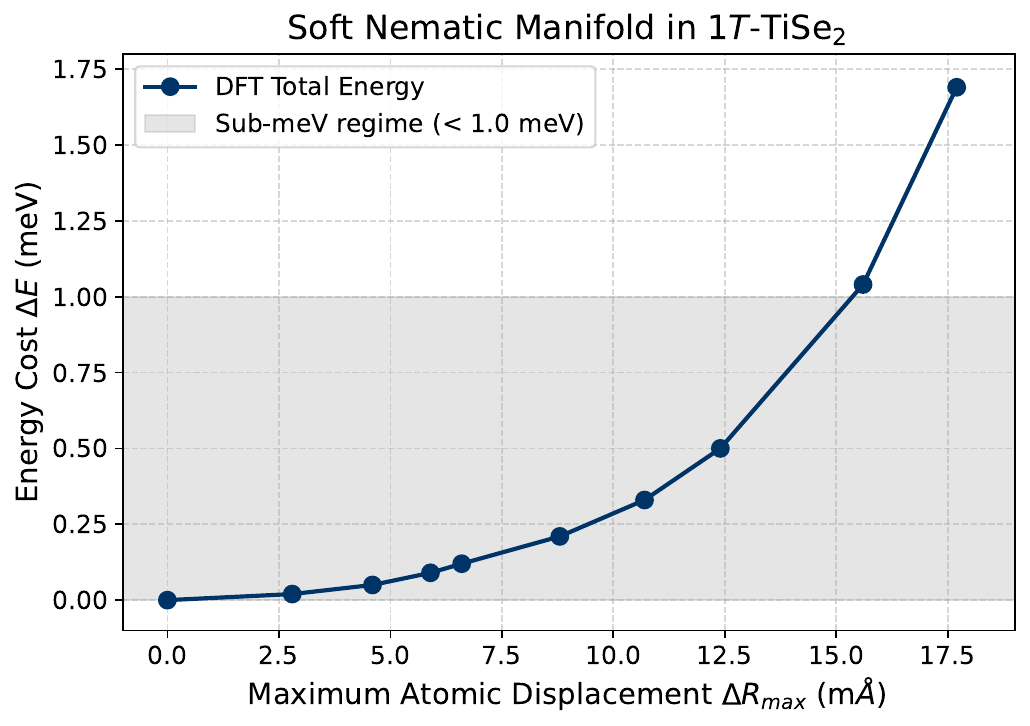}
    \caption{\textbf{Energy landscape of the soft nematic manifold.} Density functional theory calculations of the total energy cost ($\Delta E$) as a function of the maximum atomic displacement ($\Delta R_{max}$) relative to the chiral $C2$ ground state. The distortion path describes the continuous transition from the $C2$ minimum into triclinic $P1$ configurations via intra-trimer bond disproportionation. The landscape exhibits an exceptionally shallow harmonic well. The shaded gray area highlights the sub-meV fluctuation regime ($< 1.0$ meV). This broad, near-zero-energy basin allows the lattice to dynamically sample fully asymmetric configurations, reconciling the time-averaged $C2$ chirality with instantaneous $P1$ symmetry loss.\label{fig:soft_manifold}}
\end{figure}

\begin{table}[htbp]
\centering
\caption{\textbf{Symmetry-mode projection amplitudes for fully relaxed CDW structures.} Amplitudes for the atomic displacement modes along the three propagation vectors ($\mathbf{q}_1, \mathbf{q}_2, \mathbf{q}_3$), extracted from fully relaxed DFT coordinates. Values are presented in milli-Ångströms (m\AA), converted from fractional units using the nominal in-plane lattice parameter $a = 7.075$ \AA.\label{tabS9}}
\small\setlength{\tabcolsep}{4pt}\begin{tabular}{lcccc}
\toprule
Phase & Space Group & $\Gamma_7^3$ (L-mode, Ti) & $\Gamma_5^3$ (M-mode, Ti) & $\Gamma_4^3$ (Charge, Se) \\
\midrule
Achiral ($3Q_L$) & $P\bar{3}c1$ (\#165) & $(30.4, -30.4, -30.4)$ & $(0.0, 0.0, 0.0)$ & $(0.0, 0.0, 0.0)$ \\
Chiral ($2Q_L + 1Q_M$) & $C2$ (\#5) & $(0.0, -31.5, -31.6)^{\text{a}}$ & $(29.5, 0.0, 0.0)$ & $(0.0, 0.6, 0.6)$ \\
\bottomrule
\end{tabular}
\par
\vspace{1ex}
\small $^{\text{a}}$The slight difference between the secondary axes (31.5 vs 31.6 m\AA) reflects the finite force-convergence tolerance ($10^{-3}$ eV/\AA) of the DFT structural relaxation; an ideal $C2$ symmetry dictates identical amplitudes.
\end{table}

\section{Landau Free Energy Expansion}\label{secS3}

\subsection{Symmetry-Adapted Functional Form}

Following the symmetry analysis of the $P\bar{3}m1$ parent phase~\cite{Kim2024}, we construct a Landau free energy expansion in terms of the primary $L$-point ($\Gamma_7^3$) and $M$-point ($\Gamma_5^3$) phonon order parameters, denoted $\mathbf{d}_L$ and $\mathbf{d}_M$, respectively. The functional is restricted to lowest-order invariants permitted by momentum conservation and the $D_{3d}$ point group.

The total free energy is decomposed into single-mode sectors and an interaction term:
\begin{equation} \label{eq:FL_Total}
F(\mathbf{d}_L, \mathbf{d}_M) = F_L(\mathbf{d}_L) + F_M(\mathbf{d}_M) + F_{int}(\mathbf{d}_L, \mathbf{d}_M).
\end{equation}
The single-mode sectors encapsulate the harmonic drive, anharmonic stabilization, and intra-component repulsion:
\begin{equation} \label{eq:FL}
F_L = a_L \sum_{i} d_{Li}^2 + b_L \sum_{i} d_{Li}^4 + \lambda_L \sum_{i < j} d_{Li}^2 d_{Lj}^2,
\end{equation}
\begin{equation} \label{eq:FM}
F_M = a_M \sum_{i} d_{Mi}^2 + b_M \sum_{i} d_{Mi}^4 + \lambda_M \sum_{i < j} d_{Mi}^2 d_{Mj}^2,
\end{equation}
where $i=1,2,3$ indexes the three propagation vectors constituting the star of $L$ and $M$. A negative quadratic coefficient ($a < 0$) indicates a harmonic instability, while a positive quartic coefficient ($b > 0$) ensures saturation. Crucially, positive $\lambda$ represents the repulsive interaction between different propagation vectors of the same branch, penalizing simultaneous condensation (e.g., the 3$Q$ state relative to 1$Q$ states).

The interaction sector allows two distinct fourth-order invariants coupling the $L$ and $M$ amplitudes:
\begin{equation} \label{eq:Fint}
F_{int} = \kappa \sum_{i} (d_{Li}^2 d_{Mi}^2) + \gamma \sum_{i \neq j} (d_{Li}^2 d_{Mj}^2).
\end{equation}
Here, $\kappa$ governs the ``on-site'' energetic penalty between condensates sharing the same propagation vector, while $\gamma$ governs the ``cross-site'' coupling between different propagation vectors. Note that trilinear couplings of the form $d_L^3$ or $d_L^2 d_M$ are forbidden by the momentum conservation laws associated with the $P\bar{3}m1$ Brillouin zone topology, leaving these quartic terms as the leading-order interaction.

\subsection{Linear Regression of DFT Energetics}

To determine the phenomenological coefficients in Equations (\ref{eq:FL}--\ref{eq:Fint}), we performed a least-squares linear regression against the DFT total energies of the structural phases listed in Table~\ref{tab:energetics} (main text) and their neighboring distortion pathways. The amplitudes $\mathbf{d}_L$ and $\mathbf{d}_M$ were extracted directly from the atomic displacement vectors relative to the $P\bar{3}m1$ parent phase. The regression yields an $R^2$ coefficient of determination of 0.994, confirming that the inclusion of intra-component coupling terms ($\lambda$) is necessary to capture the potential energy landscape quantitatively.

For the extraction of phenomenological coefficients, the dimensionless fractional mode amplitudes were mapped to Cartesian physical displacements (\AA) using the nominal in-plane lattice parameter $a = 7.075$ \AA. The minor secondary strains associated with the fully relaxed unit cells ($\Delta a/a < 0.1\%$) are implicitly integrated into the effective quartic coefficients.

The extracted coefficients are:
\begin{equation}
\begin{aligned}
a_L &= -1.18 \times 10^4 \ \text{meV}\cdot\text{\AA}^{-2}, \quad & b_L  &= 3.24 \times 10^6 \ \text{meV}\cdot\text{\AA}^{-4}, \quad & \lambda_L &= 1.38 \times 10^6 \ \text{meV}\cdot\text{\AA}^{-4}, \\
a_M &= -1.00 \times 10^4 \ \text{meV}\cdot\text{\AA}^{-2}, \quad & b_M  &= 2.58 \times 10^6 \ \text{meV}\cdot\text{\AA}^{-4}, \quad & \lambda_M &= 1.14 \times 10^6 \ \text{meV}\cdot\text{\AA}^{-4}, \\
\kappa   &= 1.82 \times 10^7 \ \text{meV}\cdot\text{\AA}^{-4}, \quad & \gamma   &= 1.46 \times 10^6 \ \text{meV}\cdot\text{\AA}^{-4}.
\end{aligned}
\end{equation}
Here, $a$ and $b$ denote the quadratic (harmonic) and quartic (anharmonic) coefficients, while $\lambda$, $\kappa$, and $\gamma$ characterize the various fourth-order interactions.

The negative quadratic coefficients ($a_L, a_M < 0$) confirm the dual-phonon instability, while the positive quartic coefficients ($b_L, b_M > 0$) ensure saturation of the condensate amplitude. The intra-component couplings ($\lambda_L, \lambda_M > 0$) are comparable in magnitude to the single-mode stiffness terms ($b$). This repulsion renormalizes the effective stiffness in the 3$Q$ limit, suppressing the amplitude of the condensate and preventing the energy of the 3$Q$ phase from being a simple linear sum of its 1$Q$ components.

The cross-branch interaction $\kappa$ ($\sim 1.8 \times 10^7 \ \text{meV}\cdot\text{\AA}^{-4}$) is significantly larger than the single-mode stiffness ($b \sim 3.2 \times 10^6 \ \text{meV}\cdot\text{\AA}^{-4}$), indicating strong repulsion between $L$ and $M$ distortions sharing the same propagation axis. Physically, this reflects the incompatibility of the anti-phase trimerization ($L$) and uniform interlayer shear ($M$) displacements acting on the same crystallographic chain. To minimize the large penalty imposed by $\kappa$ (which forbids $L+M$ mixing on single axes), the system partitions the order parameters onto distinct axes. This selective condensation stabilizes the $2Q_L + 1Q_M$ configuration, which avoids the penalties associated with simultaneous occupation of any single propagation vector by both modes.

\subsection{Effective Free Energy Surface}

To visualize the topological structure of the ground state, we parameterize the free energy landscape using collective coordinates derived from the vector order parameters. We define:

\begin{itemize}
    \item \textbf{Chiral Mixing Angle ($\theta$):} Rotation of the order parameter on the primary propagation axis $\mathbf{q}_1$. This transforms the condensate on this axis from the $L$-mode to the $M$-mode:
    \begin{equation}
        d_{L1} = \eta_0 \cos\theta, \quad d_{M1} = \eta_0 \sin\theta.
    \end{equation}
    The condensate amplitudes on the remaining two propagation axes ($d_{L2}, d_{L3}$) are held fixed at their $L$-mode values. Consequently, $\theta=0$ corresponds to the canonical $P\bar{3}c1$ ($3Q_L$) configuration, while $\theta=\pi/2$ corresponds to the onset of $C_2$ chirality ($2Q_L + 1Q_M$).
    \item \textbf{Nematic Amplitude Imbalance ($\delta$):} Relative amplitude difference between the secondary axes:
    \begin{equation}
    d_{L2} = \eta_0(1+\delta), \quad d_{L3} = \eta_0(1-\delta).
    \end{equation}
    This coordinate quantifies the bond disproportionation degree of rotational symmetry breaking orthogonal to the primary chiral axis.
\end{itemize}

Substituting these coordinate constraints into the full Landau expansion (Equations~\ref{eq:FL_Total}--\ref{eq:Fint}) yields the effective 2D energy surface. Note that because this specific pathway involves $M$-mode condensation on only one axis ($i=1$), the intra-component repulsion $\lambda_M$ vanishes identically ($d_{Mj} = 0$ for $j \neq 1$). However, the $L$-mode repulsion $\lambda_L$ acts between all three axes. Collecting terms by their trigonometric and polynomial dependence on $\theta$ and $\delta$, the energy surface takes the form:

\begin{equation}
F_{\text{eff}}(\theta, \delta) = C + B_\theta \sin^2\theta + A_\theta \sin^4\theta + B_\delta \delta^2 + A_\delta \delta^4 + C_{mix} \delta^2 \sin^2\theta.
\end{equation}

The effective coefficients $(B_\theta, A_\theta, \dots)$ are algebraically coupled to the primary Landau parameters $(a_i, b_i, \lambda_i, \kappa, \gamma)$ through the defined transformation. Introducing the scaled amplitudes $A_i = a_i \eta_0^2$, $B_i = b_i \eta_0^4$, $\Lambda_L = \lambda_L \eta_0^4$, $K = \kappa \eta_0^4$, and $\Gamma = \gamma \eta_0^4$, where $\eta_0 \approx 0.0315$~\AA{} represents the physical equilibrium amplitude of the primary $L$-mode distortion derived from DFT. The explicit mappings are derived:
\begin{equation} \label{eq:mappings}
\begin{aligned}
C &= 3A_L + 3B_L + 3\Lambda_L \\
B_\theta &= A_M + K + 2\Gamma - A_L - 2B_L - 2\Lambda_L \\
A_\theta &= B_L + B_M - K \\
B_\delta &= 2A_L + 12B_L \\
A_\delta &= 2B_L + \Lambda_L \\
C_{mix} &= 2\Gamma - 2\Lambda_L
\end{aligned}
\end{equation}

First, consider the chiral rotational coordinate $\theta$. From the mapping relations (Equation~\ref{eq:mappings}) and the fitted regression parameters, the effective coefficients are evaluated as  $A_\theta \approx -12.2$ meV and $B_\theta \approx 13.5$ meV. The negative quartic coefficient ($A_\theta < 0$), governed primarily by the large cross-mode repulsion $K$, ensures an anharmonic drive toward the phase boundary. The positive quadratic curvature ($B_\theta > 0$) establishes the local metastability of the initial $P\bar{3}c1$ configuration ($\theta=0$). In this rigid-amplitude parameterization ($x = \sin^2\theta$), the polynomial maximum occurs at $x_{max} \approx 0.55$, corresponding to the energetic saddle point along the transition path. While the unrelaxed rigid boundary ($x=1$) sits higher in energy, subsequent amplitude relaxation ($\eta_M \neq \eta_L$) drops the $C_2$ ($\theta=\pi/2$) state well below the $P\bar{3}c1$ energy, establishing it as the absolute global minimum. This topology confirms that the strong $L+M$ on-site repulsion ($K$) creates the distinct double-well potential that fundamentally separates the chiral and achiral phases.

Second, regarding the nematic coordinate $\delta$, the effective harmonic stiffness is governed directly by $B_\delta = 2A_L + 12B_L$. Using the physical amplitude scale, we evaluate this topological stiffness as $B_\delta \approx 14.9$ meV. This positive curvature guarantees that the $C_2$ phase is a stable minimum rather than a spontaneous instability. Furthermore, the cross-coupling coefficient $C_{mix} = 2(\Gamma - \Lambda_L) \approx 0.16$ meV is effectively negligible, indicating that this nematic softness is almost entirely decoupled from the chiral rotation. We note that this analytical projection accurately captures the physical scale of the nematic softness without requiring higher-order corrections: evaluating the rigid polynomial at the extreme limit $\delta = 0.31$ yields an energetic penalty of $\sim$1.5 meV, which is in agreement with the 1.69 meV derived directly from fully relaxed DFT calculations (Table~\ref{tabS8}). This quantitative alignment dictates the uniquely shallow topology of the nematic manifold, creating a continuously accessible, sub-meV energy basin for triclinic distortions that unifies the thermodynamic stability of the chiral phase with the dynamic symmetry lowering observed spectroscopically.

\section{Quantum fluctuation along the nematic collective coordinate}
\label{secS4}

\subsection{Construction of the potential and the effective mass}

The potential along the nematic coordinate $\delta$ was generated starting
from the fully triclinic ($\delta = 0.31$, $P1$) structure: an
unconstrained structural relaxation of this configuration proceeds
continuously back to the $C2$ minimum, with the total energy and the
maximum force decreasing monotonically. Single-point energy evaluations at
intermediate geometries of this relaxation trajectory yield the values
compiled in Table~\ref{tabS8}. No intervening local minimum is encountered along
the path, confirming the single-well topology of the manifold. Since the
twofold rotation of $C2$ exchanges the two active $L$-mode axes
($\delta \rightarrow -\delta$), only the $\delta \geq 0$ branch was
computed explicitly; the negative branch is its exact symmetry image and
the fitted even polynomial enforces this symmetry identically.

An even polynomial $V(\delta) = a\delta^2 + b\delta^4$ was fitted to the
data points whose energy difference from the minimum exceeds the numerical
noise floor of the calculations ($\approx 0.05$~meV per $2\times2\times2$
supercell). The adopted coefficients are listed in Table~\ref{tabS10}; varying the
included points changes $a$ by $\lesssim 10\%$. The fitted curvature is
compatible with the independent estimate from the Landau analysis of
Sec.~\ref{secS3} ($B_\delta \approx 14.9$~meV).

The effective mass of the collective coordinate was extracted from the
displacement field between adjacent configurations along the relaxation
trajectory,
\begin{equation}
M_{\mathrm{eff}} = \sum_{\alpha} m_\alpha
\left| \frac{d\mathbf{R}_\alpha}{d\delta} \right|^2 ,
\label{eq:Meff}
\end{equation}
evaluated numerically as the average over consecutive trajectory pairs
(pair-to-pair variation $\lesssim 25\%$), giving
$M_{\mathrm{eff}} = 3.6 \times 10^{-47}$~kg\,m$^2$.

\subsection{Ground state of the one-dimensional coordinate}

\begin{figure}[htbp]
    \centering
    \includegraphics[width=0.8\textwidth]{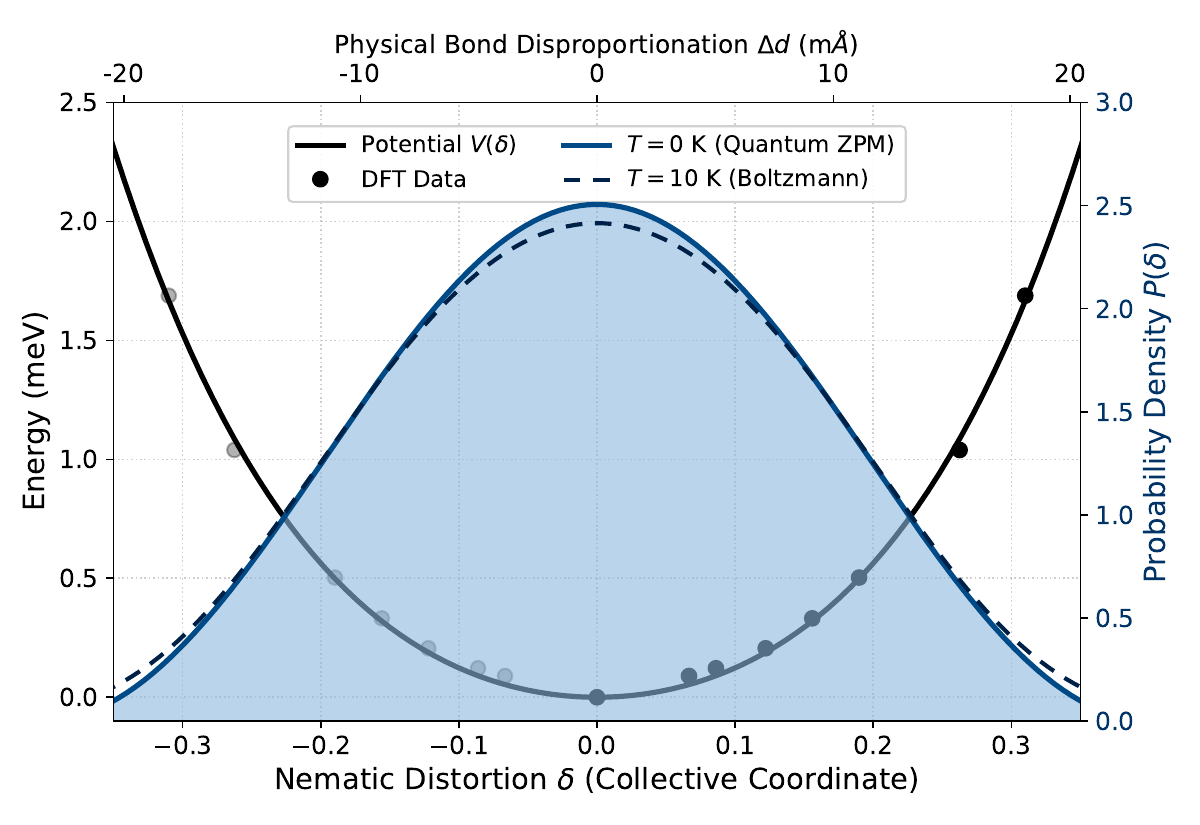}
    \caption{Quantized nematic coordinate on the soft manifold. Black
    curve: fitted potential $V(\delta) = a\delta^2 + b\delta^4$
    ($a = 11.8$, $b = 57.9$ meV per supercell), fitted to the DFT total
    energies (filled circles) above the numerical-noise floor; hollow
    circles are the symmetry images at negative $\delta$. Blue curve:
    ground-state probability density $|\psi_0(\delta)|^2$ ($T = 0$)
    obtained from the one-dimensional Schr\''odinger equation with
    $M_{\mathrm{eff}} = 3.6\times10^{-47}$ kg\,m$^2$ ($E_0 = 5.4$ meV;
    right axis). Dashed curve: classical Boltzmann distribution at
    $T = 10$ K. The ground state is exactly symmetric
    ($\langle\delta\rangle = 0$) while spanning
    $\sqrt{\langle\delta^2\rangle} = 0.26$ (top axis: approximate
    corresponding intra-trimer bond disproportionation). The potential
    beyond $|\delta| = 0.31$ is the quartic continuation of the
    fit.\label{figS3}}
\end{figure}

The one-dimensional Schr\``odinger equation for $\delta$ was solved on a uniform grid spanning $[-0.8, 0.8]$ with hard-wall boundaries; the ground-state density is negligible well inside the box, rendering the results insensitive to the boundary. The ground-state energy is $E_0 = 5.4$~meV, and no excited-state properties enter the conclusions of the present work. The probability density $|\psi_0(\delta)|^2$ (Figure~\ref{figS3}) is exactly symmetric, $\langle\delta\rangle = 0$, with $\sqrt{\langle\delta^2\rangle} = 0.26$. Using the proportionality
between $\delta$ and the intra-trimer bond disproportionation ($\sim$58 m\AA{} per unit $\delta$; Table~\ref{tabS8}), the zero-point wavefunction spans
configurations with an RMS disproportionation of $\sim$15~m\AA{} already at $T = 0$. The classical Boltzmann distribution at 10~K is slightly
narrower ($\sqrt{\langle\delta^2\rangle} \approx 0.19$), showing that at low temperature the broadening is dominated by quantum zero-point motion
rather than thermal population. Notably, the assumed corrugation scale of the well, $\sim$1~meV $\approx k_B \times 10$~K, is reached only at
$|\delta| \approx 0.25$, practically coinciding with the width of the zero-point wavefunction itself.

\begin{table}[htbp]
\centering
\caption{Quantities of the one-dimensional quantization of the nematic
coordinate $\delta$.\label{tabS10}}
\begin{tabular}{ll}
\toprule
Quantity & Value \\
\midrule
Fit coefficients (meV/supercell) & $a = 11.8$, $b = 57.9$ \\
Effective mass $M_{\mathrm{eff}}$ & $3.6\times10^{-47}$ kg\,m$^2$ \\
Characteristic mode energy $\hbar\omega$ & $\approx 6.7$ meV ($\approx 78$ K) \\
Ground-state energy $E_0$ & 5.4 meV \\
$\langle\delta\rangle$ & 0 (exact, by symmetry) \\
$\sqrt{\langle\delta^2\rangle}$ at $T = 0$ & 0.26 ($\sim$15 m\AA{} RMS $\Delta d$) \\
$\sqrt{\langle\delta^2\rangle}$ at 10 K (Boltzmann) & 0.19 (classical reference) \\
\bottomrule
\end{tabular}
\end{table}

\subsection{Limitations}

The model treats the local $\delta$ excursion of a single $2\times2\times2$
supercell within the fixed chiral ($\theta = \pi/2$) sector; it does not
describe the chirality-reversal path across the $P\bar{3}c1$ saddle, which
requires correlated motion over many supercells. The mutual coupling of the
two coordinates is negligible within the fitted Landau functional
($C_{\mathrm{mix}} \approx 0.16$~meV; Sec.~\ref{secS3}). The sampled energies follow
the relaxation trajectory, which closely approximates the minimum-energy
path along this softest structural coordinate. Beyond $|\delta| = 0.31$
the potential is the quartic continuation of the fit; ground-state
expectation values vary by $\lesssim 15\%$ under alternative smooth
extrapolations.

\section{Band Unfolding and ARPES Simulation}

\begin{figure}[htbp]
\centering
\includegraphics[width=0.9\textwidth]{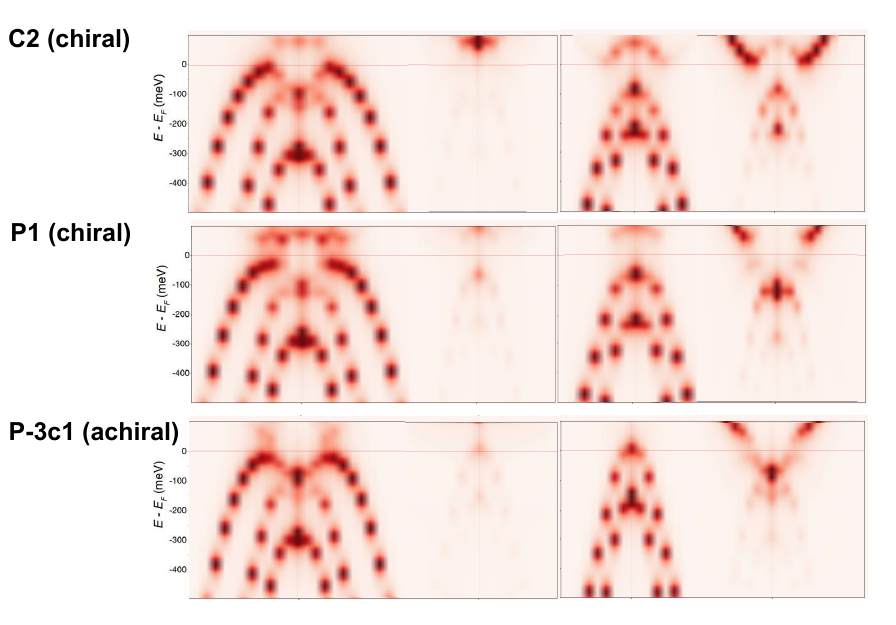}
\caption{\textbf{Simulated ARPES spectral functions.}
\textbf{(Top)} Simulated spectral weight for the chiral $C2$ ($2Q_L+1Q_M$) phase.
\textbf{(Middle)} Simulated spectral weight for the chiral $P1$ ($\delta \sim 0.3$) phase.
\textbf{(Bottom)} Simulated spectral weight for the achiral $P\bar{3}c1$ ($3Q_L$) phase.\label{fig:arpes}}
\end{figure}

\end{document}